\documentclass[twocolumn,aps,prl,showpacs]{revtex4}

\usepackage{epsfig}
\usepackage{graphics}

\newcommand{\ha}{\hat{a}}

\newcommand{\hc}{\hat{c}}

\begin{document}


\title{Long-distance teleportation of atomic qubits via optical interferometry}



\author{Y. P.  Huang}
\author{M. G. Moore}
\affiliation{Department of Physics \& Astronomy, Ohio University, Athens, OH  45701\\
		Department of Physics and Astronomy, Michigan State University, East Lansing, MI 48824}


\date{\today}

\begin{abstract}
The problem of long-distance teleportation of single-atom qubits via a common photonic channel is examined within the framework of a Mach-Zender optical interferometer.  As expected, when a coherent state is used as input,  a high-finesse optical cavity is required to overcome sensitivity to spontaneous emission. 
However, we find that  a number-squeezed light field in a twin-Fock state can in principle create useful entanglement without cavity-enhancement. Both approaches require single photon counting detectors, and best results are obtained by combining cavity-feedback with twin-fock inputs. Such an approach may allow a fidelity of $.99$ using a two-photon input and currently available mirror and detector technology. In addition, the present
approach can be conveniently extended to generate multi-site entanglement and entanglement swapping, both of which
are necessities in quantum networks.
\end{abstract}

\maketitle
Transferring quantum data between quantum processors and quantum memories, and/or  between distant quantum devices requires a protocol for high-fidelity, deterministic quantum teleportation. Quantum teleportation \cite{BenBraCre93,BouPanMat97} requires both quantum and classical channels linking the source and target qubits. The quantum channel is either photonic, utilizing entangled photon-pairs \cite{BouPanMat97,MarRieTit03} or cavity-decay photons \cite{BosKniPle99,ZheGuo00,DuaKim03,DiMutScu05} or atomic, as in recent trapped-ion experiments \cite{RieHafRoo04,BarChiSch04}.  For long-distance and/or high-speed teleportation, a photonic quantum channel is clearly ideal, as photons are robust carriers of quantum information that travel at the speed of light. As isolated trapped-atomic qubits have long coherence times and are easily manipulated with electromagnetic fields, it is of general interest to consider the problem of creating entanglement between two isolated atomic qubits via their mutual interaction with a single photonic channel. The primary obstacle to such a protocol lies in the problem of eliminating spontaneous emission while obtaining a sufficiently strong atom-photon interaction. Experimentally, this problem has been overcome only by use of collective states of atomic ensembles \cite{DuaCirZol00,JulKozPol01,MatKuz04}.

In this Letter we investigate an approach in which single-atom qubits are entangled by use of an optical Mach-Zender interferometer, thus avoiding collisional decoherence mechanisms inherent in atomic ensembles. It is well-known that focussed laser pulses interact only weakly with single-atoms, and that any quantum information imprinted on the beam by an atom is effectively destroyed by spontaneous emission \cite{EnkKim01}. Our goal is to use
the extreme sensitivity of the interferometer to detect for the weak atom-photon interaction. In order to avoid spontaneous emission we consider both the standard approach of high-finesse optical resonators\cite{HooKimYe01, RaiBruHar01}, and/or using number-squeezed photon input states to increase the interferometer sensitivity \cite{YurMcCKla86,HolBur93,PezSme06,HeiHorRey87,LanFriRon02}. We find that both approaches require detectors with single-photon resolution \cite{IrvHenBou06}. While this may appear daunting, we find that teleportation with a fidelity of $f=.99$ should be possible using only two photons and a cavity which cycles a photon $10^4$ times. This requires a single photon-on-demand \cite{KuhHenRem02} injected into each interferometer input, with an accurate measurement of the two-photon output state, which appears within the realm of experimental feasibility.

In our scheme,  a single pulse of light is passed through a Mach-Zender interferometer, with the different 'arms' of the interferometer corresponding to different photon polarization states. The beam passes through two atomic qubits, i.e. trapped ions, neutral atoms and/or quantum dots, such that each polarization state interacts with a different internal atomic state. This can be achieved using an 'X'-type scheme, as described in \cite{DuaCirZol00}, in which the Zeeman sublevels of an $F=1/2$ ground state form the qubit, or in a $\Lambda$-type level scheme, with the $m=\pm1$ states of an $F=1$ ground state forming the qubits. In both cases, the 'arms' of the interferometer would correspond to orthogonal circular polarization states.  The interferometer output is determined by a state-dependent phase acquired via the atom-photon interaction. This requires a large detuning from the atomic resonance, as there is no phase acquired on resonance. Measurement of a phase imbalance cannot determine which qubit contributed the phase-shift, resulting in entanglement from which teleportation can be achieved. We envision generalizing such a device to a complete set of quantum computation protocols whereby stationary single-atom qubits are held in isolated traps, with arbitrary single-atom and multi-atom operations achieved via sequences of light pulses guided amongst the atoms and into detectors by fast optical switching. The goal of this paper is to perform a theoretical analysis of interferometric teleportation, and to determine the fundamental limitations imposed by quantum mechanics.

\begin{figure}
    \includegraphics[height=33mm]{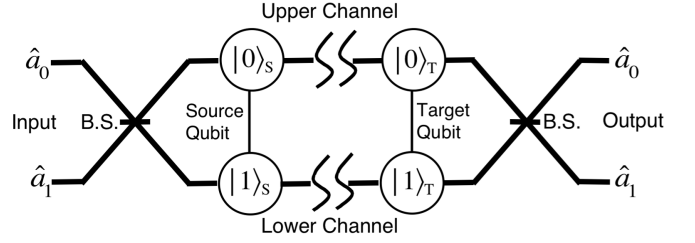}
    \caption{Schematic setup of teleportation with optical Mach-Zehnder interferometer. The interferometer is consisted of two identical $50/50$ beamsplitters, where each transmitted photon gains a phase shift of $\pi/2$.
 \label{setup}}
 \end{figure}

We consider an arbitrary atomic qubit based on
two degenerate hyperfine states, labeled as  $|0\rangle$ and $|1\rangle$. The goal 
is to map the initial quantum state of a source qubit, $|\psi_i\rangle_S=c_0
|0\rangle_S+c_1 |1\rangle_S $ onto the target qubit $|\psi\rangle_T$, prepared initially in the state $|\psi_i\rangle_T=(|0\rangle_T+|1\rangle_T)/\sqrt{2}$.
The qubits are placed inside a Mach-Zehnder interferometer
with the setup depicted in Fig.\ref{setup}, i.e. the states $|0\rangle_S$ and $|0\rangle_T$ interact with photons in the upper arm of the interferometer, while $|1\rangle_S$ and $|1\rangle_T$ interact with the lower arm.  The interferometer output is 
determined by the phase-shift acquired via the atom-photon interaction. The states $|01\rangle_{ST}$ and $|10\rangle_{ST}$ both result in zero net phase-shift, while the states $|00\rangle_{ST}$ and $|11\rangle_{ST}$ have equal and opposite non-zero phase-shift. Measuring the photon number distribution at the interferometer output collapses the qubits onto an entangled Bell-state, from which teleportation can be achieved via single qubit rotations and a source-qubit measurement. 

The basic set-up for entanglement generation is as follows. For the coherent-state input, the upper channel, described by annihilation operator $\ha_0$, is initially in a coherent state, while the lower channel $\ha_1$ is in the vacuum state. A detector is used to count the photons coming from the upper
output channel, while output in the lower channel is
unmeasured. A null result, meaning zero photons detected, results in the qubits collapsing onto a joint state $c_0
|01\rangle_{ST}+c_1 |10\rangle_{ST}+|\epsilon\rangle$, where $|\epsilon\rangle$ is the intrinsic quantum error due to the possibility of a false null result.  This error, sets the upper limit of the obtainable teleportation fidelity.
If $n\neq 0$ photons are detected, the qubits will collapse to 
$c_0 |00\rangle_{ST}+c_1 (-1)^n |11\rangle_{ST}$, without intrinsic sensitivity error.  The dual-Fock input set-up differs in that the photon number-difference between the outputs must be measured. In this case case, a result of zero number-difference constitutes a null result.

To derive these results, we construct the propagator for the passage of photons through the interferometer,   $\hat{U}=\hat{U}_{BS}\hat{U}_T\hat{U}_S\hat{U}_{BS}$, where $\hat{U}_{BS}$ is the beam-splitter propagator, and $\hat{U}_S$ and $\hat{U}_T$ are the propagators for the interaction with the source and target qubits, respectively. The beam-splitter propagator is $\hat{U}_{BS}=\exp[-i(\ha_o^\dag\ha_1+\ha_1^\dag\ha_0)\pi/4]$, while the qubit-photon interaction propagators are $\hat{U}_\mu=\exp[-i\theta(\ha_0^\dag\ha_0\hc_{\mu0}^\dag\hc_{\mu0}+\ha_1^\dag\ha_1\hc^\dag_{\mu 1}\hc_{\mu 1})]$, where  $\hat{c}_{\mu\nu}$ is the annihilation operator for an atom at location $\mu\in\{S,T\}$ in internal state $\nu\in\{0,1\}$. This interaction operator is valid in the far-off-resonance regime, where the electronically excited state can be adiabatically eliminated. The interaction is governed by the phase-shift $\theta=|d{\cal E}(\omega)|^2\tau/(\hbar^2\Delta)$, where $\tau$ is the atom-photon interaction time, $\Delta$ is the detuning between the laser and atomic resonance frequencies,  $d$ is the electric dipole moment and ${\cal E}(\omega)=\sqrt{\hbar\omega/(2\epsilon_0V)}$ is the `electric field per photon' for laser frequency $\omega$ and mode-volume $V$. Introducing the spontaneous emission rate $\Gamma=d^2\omega^3/(3\pi\epsilon_0\hbar c^3)$, taking the photon mode as having length $L$ and width $W$ (at the location of the atom), and taking the interaction time as $\tau=L/c$, we arrive at the single-atom phase-shift
\begin{equation}
\label{theta}
    \theta=(3/8\pi)(\lambda/W)^2(\Gamma/\Delta),
\end{equation}
where $\lambda$ is the laser wavelength. This is the phase-shift acquired by an off-resonant photon forward-scattered by a single atom, and is independent of the pulse length. 

Treating first the coherent state input, the initial
state of the complete system can be expressed as
$|\Psi_i\rangle=e^{-\alpha\ha^{\dagger}_0+\alpha^\ast \ha_0}|\Psi_0\rangle$, where $|\Psi_0\rangle=|0\rangle\otimes
(1/\sqrt{2})(c_0|00\rangle_{ST}+c_0|01\rangle_{ST}+c_1|10\rangle_{ST}+c_1|11\rangle_{ST})$,
with $|0\rangle$ being the electromagnetic vacuum state. The state of the system at the interferometer output is then given by
 $|\Psi_f\rangle=\hat{U}|\Psi_i\rangle=e^{-\alpha
\hat{U}\ha^{\dagger}_0\hat{U}^\dagger+\alpha^\ast
\hat{U}\ha_0\hat{U}^\dagger}|\Psi_0\rangle$.  Introducing
$\sigma_z=\frac{1}{2}\sum_{\mu=S, T} (\hc^{\dagger}_{\mu 0}\hc_{\mu
0}-\hc^{\dagger}_{\mu 1}\hc_{\mu 1})$, we find that $\hat{U}\ha^{\dagger}_0\hat{U}^\dagger=-i e^{-i\theta}[\sin (\theta
\sigma_z) \ha_0+\cos(\theta\sigma_z) \ha_1]$. Thus
$|\Psi_f\rangle$ can be written as
\begin{equation}
\label{output}
    |\Psi_f\rangle = \sum_{i,j=0,1} c_{i}|\bar{\alpha}\sin\theta_{ij},
    \bar{\alpha}\cos\theta_{ij}\rangle\otimes|ij\rangle_{ST}
\end{equation}
where $\theta_{ij}=\theta\, (1-i-j)$ and $\bar{\alpha}=-i\alpha e^{i\theta}$, and the state $|\alpha_0,\alpha_1\rangle$ indicates a two-mode coherent state for the two interferometer outputs. Expanding the upper channel onto photon number-eigenstates and making the small-angle approximation gives $|\Psi_F\rangle=\sum_{n=0}^\infty|n\rangle_0\otimes|\bar{\alpha}\rangle_1\otimes|\phi_n\rangle_{ST}$, where $|n\rangle_0$ indicates a state with $n$ photons in the upper output, $|\bar{\alpha}\rangle_1$ indicates a coherent-state in the lower output, and
\begin{eqnarray}
\label{coherentoutput}
    |\phi_0\rangle_{ST}&=&c_0|01\rangle+c_1|10\rangle+|\epsilon\rangle\\
    |\phi_{n\neq 0}\rangle_{ST}&=&f_n\left[c_0|00\rangle + (-1)^nc_1|11\rangle\right],
\end{eqnarray}
where $|\epsilon\rangle=e^{-|\alpha|^2\theta^2/2}(c_0|00\rangle_{ST}+c_1|11\rangle_{ST})$ and $f_n=(\bar{\alpha}^n/\sqrt{n!}) e^{-|\alpha|^2\theta^2/2}$.

The photon number in the upper channel is then measured with single-photon resolution. The probability of detecting $n$ photons $P(n)$ is given by
$P(n)=\frac{1}{2} [\delta_{n,0}+ e^{-N \theta^2} N
\theta^{2n}/n!]$, where $N=|\alpha|^2$ is the mean input photon number.
The probability of detecting zero photons is thus
$P(0)=1/2(1+\epsilon)$, with $\epsilon=e^{-N \theta^2}$. Hence the null result will be obtained $\sim50\%$ of the time, with the qubit-state will collapsing onto the state
$[c_0|01\rangle_{ST}+c_1|10\rangle_{ST}+\sqrt{\epsilon}(c_0|00\rangle_{ST}+c_1|11\rangle_{ST})]/\sqrt{1+\epsilon}$, so that the fidelity of teleporation is
$1/(1+\epsilon)\approx1-\epsilon$, with the reduction due to the possibility of a false null result.  The condition for faithful teleportation is $N\theta^2\gg 1$, characteristic of a shot-noise-limited interferometer.

The remaining 50\% of the time, a photon-number $n\neq 1$ is detected, with the qubit-state collapsing onto 
$c_0 |00\rangle+(-1)^nc_1|11\rangle]$.  The $(-1)^{n}$ term
comes from the phase difference between number-states for the coherent states $|\alpha\rangle$ and $|-\alpha\rangle$, i.e. while measuring photon number can not distinguish the states $|00\rangle$ and $|11\rangle$, it can introduce relative phase between them. If the photon number is not measured exactly, then tracing over the photon number create a statistical mixture of  $|00\rangle$ and $|11\rangle$, so that it is necessary to determine the photon number exactly. This difficulty is somewhat mitigated by the fact that the
average photon number is $\bar{n}_0=-\ln \epsilon$, i.e.
only 7 photons must be counted for a fidelity of 99.9\%. 

Once the entangled qubit pair is generated, completing the teleportation requires that the qubits be disentangled. This can be accomplished in the following manner.
Conditional upon a null
result, a $\pi$-pulse is applied to the source qubit, flipping
$|0\rangle_S \leftrightarrow |1\rangle_S$. In the case of an odd photon number $n$, a relative $\pi$ phase is applied to the state $|1\rangle_{S}$ (or $|1\rangle_{T}$). After these steps, the qubits' state becomes $c_0 |00\rangle_{ST}+c_1|11\rangle_{ST}$. 3)
a $\pi/2$-pulse is applied to the source (or the target)
qubit, transforming the state into
$[c_0|00\rangle-ic_0|10\rangle-ic_1|01\rangle+c_1|11\rangle]/\sqrt{2}$.
The state of the source qubit is measured.  If the result is $|0\rangle_S$, the target
qubit will collapse to $c_0 |0\rangle_T-i c_1 |1\rangle_T$, in which case a
$\pi/2$ phase is imprinted onto $|1\rangle_T$. If the
result is $|1\rangle_S$, a $\pi/2$ phase is
imprinted onto $|0\rangle_T$, resulting in the desired state $c_0|0\rangle_T+c_1|1\rangle_T$.

Aside from the technical challenge of single-photon counting, the fundamental quantum-mechanical
barrier to successful teleportation lies in finding a balance between phase-shift detection and spontaneous-emission avoidance, as a single spontaneously scattered photon can destroy the coherence of a qubit.
The spontaneous emission probability for a single qubit is $\theta N\Gamma/\Delta$, which becomes negligible when $\theta N \Gamma/\Delta\ll 1$. This condition must be satisfied without violating the shot-noise-sensitivity condition $N\theta^2\gg 1$. From equation (\ref{theta}) it follows that compatibility  requires $16(W/\lambda)^2\ll 1$, which clearly violates the standard optical diffraction limit. That such a scheme cannot work is in agreement with the well-known results of \cite{EnkKim01}. To overcome this difficulty, we can place the two qubits in seperate high-finesse optical cavities, with mechanical Q-switching employed to restrict the photon to $M$ passes through each qubit. This will increase the phase-shift $\theta$ and the spontaneous emission probability $P_{sp}$ by a factor of $M$. This relaxes the compatibility condition to $8(W/\lambda)^2\ll M$, which can be satisfied without sub-wavelength focussing.

The failure probabilities due to
interferometry sensitivity and spontaneous emission are
$\epsilon=e^{-NM\theta^2}$ and $P_{SP}=2NM\theta\Gamma/\Delta$, respectively. Setting $P_{sp}=\epsilon=.01$, corresponding to a teleportation fidelity of $.99$, and taking $W/\lambda=3$ gives $M=-144\log{\epsilon}/\epsilon=6.6\times 10^5$, which is large but not outside the range of current experimental techniques. For these parameters, the mean number of photons in the upper output is $\bar{n}_0=4$, and the input photon number is restricted only by the condition $N\left(\Gamma/\Delta\right)^2=144\, \epsilon/M= 4.4\times 10^{-6}$, together with the off-resonant condition $\Delta\gg\Gamma$.

To achieve a higher fidelity, and/or to eliminate the need for a high-finesse resonator, one
can employ sub-shot-noise interferometry. We now investigate the fundamental limits to teleportation fidelity when a twin-fock photon input state is used increase the interferometer sensitivity. The input state is now $|N,N\rangle$, where
$|k,l\rangle=(\ha^\dagger_0)^k(\ha^\dagger_1)^l|0\rangle/\sqrt{k!~l!}$. Following the previous approach, the output state is now
\begin{equation}
\label{df}
    |\Psi_f\rangle = \sum_{ij}c_i|ij\rangle_{ST}\otimes\sum^{N}_{m=-N} \chi_m(\theta_{ij}) |N+m, N-m\rangle ,
\end{equation}
where
\begin{eqnarray}
    \chi_m(\theta_{ij}) =\sum^{\min\{N, N-m\}}_{l=\max\{0, -m\}} (-1)^{m+l} \
    \left(\begin{array}{c}m+l \\N \end{array}\right)
    \left(\begin{array}{c}l \\N \end{array}\right) \times \nonumber \\
    \frac{\sqrt{(N+m)! (N-m)!}}{N! }\sin^{m+2l}\theta_{ij} \cos^{2N-m-2l}\theta_{ij}.
\end{eqnarray}
The desired two-qubit entangled state is then created by measuring the
photon number difference between the upper and lower outputs. It is seen from
(\ref{df}) that the probability of detecting a difference of $2m$ is
$P(2m)=\sum_{ij} |c_i|^2 \chi^2_m(\theta_{ij})$. Thus the probability to detect zero photon number difference is $P(0)=1/2(1+\eta)$, where $\eta=\chi^2_0(\theta)$ is the probability of a false null result. If the result is $m=0$, the qubits collapse onto
$[c_0|01\rangle_{ST}+c_1|10\rangle_{ST}
+\sqrt{\eta}(c_0|00\rangle_{ST}+c_1|11\rangle_{ST})]/\sqrt{1+\eta}$, so that
the teleportation fidelity is $1/(1+\eta)\approx 1-\eta$. If the result is $2m\neq 0$, the qubits
collapse onto $c_0|00\rangle_{ST}+(-1)^m c_1|11\rangle_{ST}$. Again the exact photon number difference must be measured in order to successfully disentangle the qubits.

\begin{figure}
 \includegraphics[height=50mm]{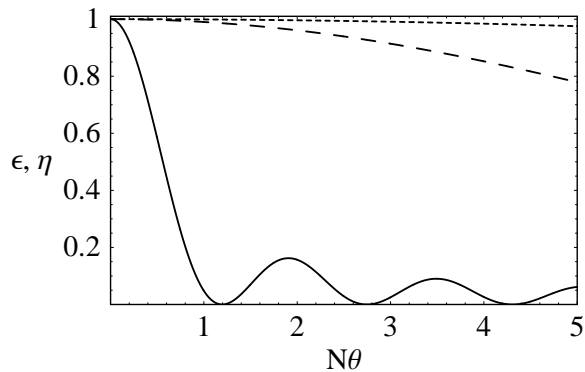}.
 \caption{Plots of intrinsic error due to interferometer sensitivity.  Error for twin-fock input, $\eta$ (solid), is compared to that from coherent-state input, $\epsilon$, with $N=100$ (dashed) and $N=1000$ (dotted) as a function of $N\theta$, respectively. Note $\eta$ is only dependent on $N\theta$. 
 \label{fig2}}
 \end{figure}

The twin-fock input thus yields results
similar to the coherent state-input,  but with the intrinsic error due to interferometer sensitivity given by $\eta$ instead of $\epsilon$. A comparison plot of $\eta$ and $\epsilon$ is shown in Fig. \ref{fig2}, in which it is seen that $\eta$
decreases with $N$ much faster than $\epsilon$. In fact, for
$N\theta < 1$, $\eta\approx e^{-N^2 \theta^2}$, which is characteristic of a Heisenberg-limited interferometer.  This means that
significantly fewer photons are required to obtain equal
fidelity, with a corresponding reduction in spontaneous emission. 
In Fig. \ref{fig2} we see that the false-null probability $\eta$ is exactly zero for a periodic set of values of $N\theta$. The first such zero occurs at $N\theta=1.196\equiv x_0$. Thus if one can precisely control $N\theta$, it is possible to achieve teleportation without intrinsic error due to false-null results. In this case, the success of teleportation is governed only by spontaneous emission probability $P_{sp}=2 N \theta \Gamma/\Delta=2 x_0\Gamma/\Delta$. The condition $N\theta=x_0$, together with (\ref{theta}), means that $\Gamma/\Delta=(16 x_0/N)\left(W/\lambda\right)^2$, so that $P_{sp}=(16 x_0^2/N)\left(W/\lambda\right)^2$. For the case of a tightly focussed beam, $W/\lambda=3$ this gives $P_{sp}=206/N$. The theoretical limit to fidelity therefore scales as $\sim1-200/N$, thus a fidelity of $f=.99$ would require $N=20,000$ atoms, while a fidelity of $f=.999$ could be achieved with $N=2\times 10^5$ atoms. An extremely high fidelity of $f=.999999$ would therefore require $2\times10^8$ atoms. The addition of a Q-switched cavity with feedback results in $P_{sp}=206/(NM$), which for $M=10^4$, would reduce the atom numbers to $N=2$ for $f=.99$, $N=20$ for $f=.999$, and $N=2\times10^4$ for $f=.999999$. We note that for the twin-Fock input, a single lost photon will disrupt the teleportation completely.

In addition to teleportation of arbitrary qubits, our scheme can be easily
generalized to generate many-qubit entanglement \cite{GreHorZei89} as well as
realize  entanglement swapping \cite{PanBouWei98,DuaLukCir01}. For example, the three-particle Greenberger-Horne-Zeilinger (GHZ)
state, $(|000\rangle_{ABC}+|111\rangle_{ABC})/\sqrt{2}$, can be created by first preparing each qubit
in the state $(|0\rangle_i+|1\rangle_i)/\sqrt{2}$, with
$i=A,B,C$. Then the two-qubit protocol is used to collapse $A$
and $B$ into the state
$(|00\rangle_{AB}+|11\rangle_{AB})/\sqrt{2}\otimes
(|0\rangle_C+|1\rangle_C)/\rangle/\sqrt{2}$. If the same two-qubit
procedure  is then applied to $B$ and $C$,  the GHZ
state is obtained. Moreover, this scheme can be extended in a straightforward manner to producing an $N$-particle cat
state. To realize
entanglement swapping, we take an initially entangled qubit
pair, $(1/\sqrt{2})(c_0|00\rangle_{AB}+c_1|11\rangle_{AB})$, and an uncorrelated qubit $(1/\sqrt{2})(|0\rangle_C+|1\rangle_C)$ and apply our protocol to qubits $B$\&$C$ to create a GHZ-like state. Then, by
disentangling $B$ in the same manner as described for the source qubit in teleportation,  one gets the swapped state $(c_0
|00\rangle_{AC}+c_1 |11\rangle_{AC})/\sqrt{2}$. 

In conclusion, we have used the formalism of the Mach-Zender interferometer to treat the problem of creating entanglement between two atoms via a common photonic channel. We have compared the results from a coherent state input with high-finesse cavity enhancement, with those from quantum-limited input states. Our results suggest that by combining the two approaches, long-distance teleportation may be achievable with current  experimental techniques. In particular, we find that a two-photon input state has a fundamental upper-limit to fidelity of $.99$, and provides the advantage that failure due to photon losses could be readily detected.




\end{document}